\documentclass{aa}

\usepackage{graphicx}
\usepackage{txfonts}
\usepackage{color}
\usepackage{stfloats}
\usepackage{amsmath}
\usepackage{multirow}

\usepackage{epsfig}
\usepackage{amsmath}
\usepackage{subfigure}
\usepackage{natbib}

\usepackage{longtable}
\usepackage{graphicx}
\usepackage{epstopdf}
\usepackage{booktabs}
\usepackage{txfonts}
\usepackage[utf8]{inputenc}%
\usepackage{graphicx}
\usepackage{txfonts}
\usepackage{stfloats}

\newcommand{\tablenotea}[1]{\parbox{17.8cm}{\indent \footnotesize{#1}}}
\newcommand{\tablenoteb}[1]{\parbox{16.3cm}{\indent \footnotesize{#1}}}
\newcommand{\tablenotec}[1]{\parbox{8.8cm}{\indent \footnotesize{#1}}}
\newcommand{\acsesc}{ACS Earth Space Chem.}

\newcommand{\cpl}{Chem. Phys. Lett.}
\newcommand{\chemrev}{Chem. Rev.}
\newcommand{\ctc}{Comput. Theor. Chem.}

\newcommand{\ijqc}{Int. J. Quantum Chem.}

\newcommand{\jacs}{J. Am. Chem. Soc.}

\newcommand{\jcompp}{J. Comput. Phys.}

\newcommand{\jpc}{J. Phys. Chem.}
\newcommand{\jpca}{J. Phys. Chem. A}

\newcommand{\mphys}{Mol. Phys.}

\newcommand{\wcms}{WIREs Comput. Mol. Sci.}

\begin{document}

\title{The chemistry of H$_2$NC in the interstellar medium and\\ the role of the C + NH$_3$ reaction\thanks{Based on observations carried out with the IRAM 30m Telescope. IRAM is supported by INSU/CNRS (France), MPG (Germany) and IGN (Spain).}}

\titlerunning{The chemistry of H$_2$NC in the interstellar medium}
\authorrunning{Ag\'undez et al.}

\author{M.~Ag\'undez\inst{1}, O.~Roncero\inst{1}, N.~Marcelino\inst{2,3}, C.~Cabezas\inst{1}, B.~Tercero\inst{2,3}, \and J.~Cernicharo\inst{1}}

\institute{
Instituto de F\'isica Fundamental, CSIC, C/ Serrano 123, 28006 Madrid, Spain\\
\email{marcelino.agundez@csic.es, octavio.roncero@csic.es} \and
Observatorio Astron\'omico Nacional, IGN, Calle Alfonso XII 3, 28014 Madrid, Spain \and
Observatorio de Yebes, IGN, Cerro de la Palera s/n, 19141 Yebes, Guadalajara, Spain
}

\date{Received; accepted}


\abstract
{We carried out an observational search for the recently discovered molecule H$_2$NC, and its more stable isomer H$_2$CN, toward eight cold dense clouds (L1544, L134N, TMC-2, Lupus-1A, L1489, TMC-1\,NH$_3$, L1498, and L1641N) and two diffuse clouds (B0415+379 and B0355+508) in an attempt to constrain its abundance in different types of interstellar regions and shed light on its formation mechanism. We detected H$_2$NC in most of the cold dense clouds targeted, 7 out of 8, while H$_2$CN was only detected in 5 out of 8 clouds. The column densities derived for both H$_2$NC and H$_2$CN are in the range 10$^{11}$-10$^{12}$ cm$^{-2}$ and the abundance ratio H$_2$NC/H$_2$CN varies between 0.51 and $>$\,2.7. The metastable isomer H$_2$NC is therefore widespread in cold dense clouds where it is present with an abundance similar to that of H$_2$CN. We did not detect either H$_2$NC or H$_2$CN in any of the two diffuse clouds targeted, which does not allow to shed light on how the chemistry of H$_2$NC and H$_2$CN varies between dense and diffuse clouds. We found that the column density of H$_2$NC is correlated with that of NH$_3$, which strongly suggests that these two molecules are chemically linked, most likely ammonia being a precursor of H$_2$NC through the C + NH$_3$ reaction. We performed electronic structure and statistical calculations which show that both H$_2$CN and H$_2$NC can be formed in the C + NH$_3$ reaction through two different channels involving two different transition states which lie very close in energy. The predicted product branching ratio H$_2$NC/H$_2$CN is very method dependent but values between 0.5 and 0.8 are the most likely ones. Therefore, both the astronomical observations and the theoretical calculations support that the reaction C + NH$_3$ is the main source of H$_2$NC in interstellar clouds.}


\maketitle

\section{Introduction}

The detection of H$_2$NC in the interstellar medium by \cite{Cabezas2021} came as a surprise. The observation of several groups of unidentified lines in the cold dense clouds L483 and B1-b motivated the search for potential carriers, among which H$_2$NC was not initially considered. Only after many plausible candidates were ruled out, H$_2$NC was found to confidently reproduce the complex spectral pattern of astronomical lines. H$_2$NC was not initially considered as a plausible carrier because it is a high-energy isomer of H$_2$CN, which lies $\sim$\,30 kcal mol$^{-1}$ above in energy \citep{Puzzarini2010,Cabezas2021}. Moreover, there are two isomers not yet detected in space which are substantially more stable than H$_2$NC. These are the trans and cis forms of HCNH, which lie $\sim$\,8 and $\sim$\,13 kcal mol$^{-1}$, respectively, above the most stable isomer H$_2$CN \citep{Puzzarini2010,Cabezas2021}. It is worth to note that although H$_2$NC is the least stable member of this family of four isomers, it is the one showing the most intense lines in L483 and B1-b.

The discovery of interstellar H$_2$NC came with several interesting implications. In addition to L483 and B1-b, H$_2$NC was also detected toward the $z$\,=\,0.89 galaxy in front of the quasar PKS\,1830$-$211, which is a high-redshift source exceptionally rich in molecules \citep{Muller2011,Tercero2020}. The fact that the abundance ratio H$_2$NC/H$_2$CN is $\sim$\,1 in L483 and B1-b and 0.27 toward PKS\,1830$-$211 suggests that this ratio behaves as the HNC/HCN ratio, with values close to one in cold dense clouds and below one in diffuse clouds \citep{Hirota1998,Sarrasin2010,Liszt2001}. Surprisingly, neither H$_2$NC nor H$_2$CN were detected toward the cyanopolyyne peak of TMC-1 (hereafter TMC-1\,CP) in spite of a previous claim of detection of H$_2$CN in this cloud by \cite{Ohishi1994}. While TMC-1\,CP is a young starless core rich in carbon chains, L483 and B1-b are more evolved sources rich in molecules such as ammonia. The detection of H$_2$NC in L483 and B1-b and its non detection in TMC-1\,CP suggests that this species is more favored in NH$_3$-rich sources. This would be in line with the suspected chemical origin of H$_2$NC due to the reaction C + NH$_3$ (see discussion in \citealt{Cabezas2021}).

In order to shed light on the issues raised by the discovery of H$_2$NC, that is, whether or not the H$_2$NC/H$_2$CN ratio behaves differently in dense and diffuse clouds, as occurs for the HNC/HCN ratio, and whether or not there is a correlation between the abundances of H$_2$NC and NH$_3$, which would reflect a chemical link between the two species, we carried out an observational survey to search for H$_2$NC, and simultaneously for H$_2$CN, toward a sample of dense and diffuse clouds with the IRAM\,30m telescope. Here we report the results of this survey. We also present a computational study of the reaction between C and NH$_3$ aimed at investigating whether or not H$_2$NC can be formed in this reaction at the low temperatures of cold interstellar clouds.

\section{Astronomical observations}

\begin{table}
\small
\caption{Sources coordinates.}
\label{table:coordinates}
\centering
\begin{tabular}{l@{\hspace{1.25cm}}r@{\hspace{1.00cm}}r}
\hline \hline
Source & RA(J2000.0) & Dec(J2000.0) \\
\hline
L1544      & 05:04:18.10 & $+$25:10:48.0 \\
L134N      & 15:54:06.55 & $-$02:52:19.1 \\
TMC-2      & 04:32:44.74 & $+$24:24:33.5 \\
Lupus-1A   & 15:42:52.40 & $-$34:07:53.5 \\
L1489      & 04:04:50.60 & $+$26:18:29.4 \\
TMC-1\,NH3 & 04:41:23.02 & $+$25:48:13.3 \\
L1498      & 04:10:51.46 & $+$25:09:58.2 \\
L1641N     & 05:36:19.01 & $-$06:22:13.3 \\
B0415+379  & 04:18:21.28 & $+$38:01:35.8 \\
B0355+508  & 03:59:29.75 & $+$50:57:50.2 \\
\hline
\end{tabular}
\end{table}

\begin{table*}[hb!]
\small
\caption{Line parameters of the strongest hyperfine components of H$_2$NC and H$_2$CN.}
\label{table:lines}
\centering
\begin{tabular}{llcrcccc}
\hline \hline
\multicolumn{1}{l}{Source} & \multicolumn{1}{c}{Molecule} & \multicolumn{1}{c}{Transition} & \multicolumn{1}{c}{$\nu_{calc}$} & \multicolumn{1}{c}{$V_{\rm LSR}$} & \multicolumn{1}{c}{$\Delta v$} & \multicolumn{1}{c}{$T_A^*$ peak} & \multicolumn{1}{c}{$\int T_A^* dv$} \\
& & & \multicolumn{1}{c}{(MHz)} &\multicolumn{1}{c}{(km s$^{-1}$)} &\multicolumn{1}{c}{(km s$^{-1}$)} & \multicolumn{1}{c}{(mK)} &\multicolumn{1}{c}{(mK km s$^{-1}$)} \\
\hline
L1544 & H$_2$NC & 1$_{0,1}$-0$_{0,0}$ $J$\,=\,1.5-0.5 $F_1$\,=\,2.5-1.5 $F$\,=\,3.5-2.5 & 72194.211 & 7.10(2) & 0.49(4) & 31.7(2) & 16.5(10) \\
L1544 & H$_2$NC & 1$_{0,1}$-0$_{0,0}$ $J$\,=\,1.5-0.5 $F_1$\,=\,2.5-1.5 $F$\,=\,2.5-1.5 & 72210.940 & 7.13(3) & 0.48(6) & 26.4(2) & 13.4(14) \\
L134N & H$_2$NC & 1$_{0,1}$-0$_{0,0}$ $J$\,=\,1.5-0.5 $F_1$\,=\,2.5-1.5 $F$\,=\,3.5-2.5 & 72194.211 & 2.40(2) & 0.47(3) & 39.4(3) & 19.6(10) \\
L134N & H$_2$NC & 1$_{0,1}$-0$_{0,0}$ $J$\,=\,1.5-0.5 $F_1$\,=\,2.5-1.5 $F$\,=\,2.5-1.5 & 72210.940 & 2.40(3) & 0.43(6) & 25.7(3) & 11.8(15) \\
TMC-2 & H$_2$NC & 1$_{0,1}$-0$_{0,0}$ $J$\,=\,1.5-0.5 $F_1$\,=\,2.5-1.5 $F$\,=\,3.5-2.5 & 72194.211 & 6.19(5) & 0.45(13) & 19.4(5) & 9.3(21) \\
TMC-2 & H$_2$NC & 1$_{0,1}$-0$_{0,0}$ $J$\,=\,1.5-0.5 $F_1$\,=\,2.5-1.5 $F$\,=\,2.5-1.5 & 72210.940 & 6.25(4) & 0.20(10) & 16.8(5) & 3.6(16)\,$^a$ \\
Lupus-1A & H$_2$NC & 1$_{0,1}$-0$_{0,0}$ $J$\,=\,1.5-0.5 $F_1$\,=\,2.5-1.5 $F$\,=\,3.5-2.5 & 72194.211 & 5.00(8) & 0.40(19) & 18.9(5) & 8.0(27) \\
Lupus-1A & H$_2$NC & 1$_{0,1}$-0$_{0,0}$ $J$\,=\,1.5-0.5 $F_1$\,=\,2.5-1.5 $F$\,=\,2.5-1.5 & 72210.940 & 5.11(5) & 0.34(10) & 17.7(5) & 6.4(17) \\
L1489 & H$_2$NC & 1$_{0,1}$-0$_{0,0}$ $J$\,=\,1.5-0.5 $F_1$\,=\,2.5-1.5 $F$\,=\,3.5-2.5 & 72194.211 & 6.66(4) & 0.50(9) & 14.1(3) & 7.4(12) \\
L1489 & H$_2$NC & 1$_{0,1}$-0$_{0,0}$ $J$\,=\,1.5-0.5 $F_1$\,=\,2.5-1.5 $F$\,=\,2.5-1.5 & 72210.940 & 6.71(5) & 0.39(10) & 14.2(3) & 5.9(14) \\
TMC-1\,NH$_3$ & H$_2$NC & 1$_{0,1}$-0$_{0,0}$ $J$\,=\,1.5-0.5 $F_1$\,=\,2.5-1.5 $F$\,=\,3.5-2.5 & 72194.211 & 5.89(3) & 0.54(8) & 22.8(2) & 13.0(17) \\
TMC-1\,NH$_3$ & H$_2$NC & 1$_{0,1}$-0$_{0,0}$ $J$\,=\,1.5-0.5 $F_1$\,=\,2.5-1.5 $F$\,=\,2.5-1.5 & 72210.940 & 5.86(4) & 0.54(9) & 22.4(2) & 12.7(18) \\
L1498 & H$_2$NC & 1$_{0,1}$-0$_{0,0}$ $J$\,=\,1.5-0.5 $F_1$\,=\,2.5-1.5 $F$\,=\,3.5-2.5 & 72194.211 & 7.71(5) & 0.57(14) & 16.7(4) & 10.2(20) \\
L1498 & H$_2$NC & 1$_{0,1}$-0$_{0,0}$ $J$\,=\,1.5-0.5 $F_1$\,=\,2.5-1.5 $F$\,=\,2.5-1.5 & 72210.940 & 7.80(2) & 0.20(10) & 20.2(4) & 4.4(13) \\
\multicolumn{8}{c}{} \\
L1544 & H$_2$CN & 1$_{0,1}$-0$_{0,0}$ $J$\,=\,1.5-0.5 $F_1$\,=\,2.5-1.5 $F$\,=\,2.5-1.5 & 73345.486 & 7.19(3) & 0.39(7) & 14.9(2) & 6.2(9) \\
L1544 & H$_2$CN & 1$_{0,1}$-0$_{0,0}$ $J$\,=\,1.5-0.5 $F_1$\,=\,2.5-1.5 $F$\,=\,3.5-2.5 & 73349.648 & 7.23(3) & 0.40(5) & 17.8(2) & 7.6(9) \\
L134N & H$_2$CN & 1$_{0,1}$-0$_{0,0}$ $J$\,=\,1.5-0.5 $F_1$\,=\,2.5-1.5 $F$\,=\,2.5-1.5 & 73345.486 & 2.53(7) & 0.40(15) & 6.6(2) & 2.8(9)\,$^a$ \\
L134N & H$_2$CN & 1$_{0,1}$-0$_{0,0}$ $J$\,=\,1.5-0.5 $F_1$\,=\,2.5-1.5 $F$\,=\,3.5-2.5 & 73349.648 & 2.45(5) & 0.50(15) & 11.8(2) & 6.3(14) \\
TMC-2 & H$_2$CN & 1$_{0,1}$-0$_{0,0}$ $J$\,=\,1.5-0.5 $F_1$\,=\,2.5-1.5 $F$\,=\,2.5-1.5 & 73345.486 & 6.40(11) & 0.67(22) & 10.5(4) & 7.5(23) \\
TMC-2 & H$_2$CN & 1$_{0,1}$-0$_{0,0}$ $J$\,=\,1.5-0.5 $F_1$\,=\,2.5-1.5 $F$\,=\,3.5-2.5 & 73349.648 & 6.47(7) & 0.46(20) & 17.0(4) & 8.2(25) \\
Lupus-1A & H$_2$CN & 1$_{0,1}$-0$_{0,0}$ $J$\,=\,1.5-0.5 $F_1$\,=\,2.5-1.5 $F$\,=\,2.5-1.5 & 73345.486 & 4.97(9) & 0.31(16) & 9.9(4) & 3.3(16)\,$^a$ \\
Lupus-1A & H$_2$CN & 1$_{0,1}$-0$_{0,0}$ $J$\,=\,1.5-0.5 $F_1$\,=\,2.5-1.5 $F$\,=\,3.5-2.5 & 73349.648 & 5.10(2) & 0.20(10) & 28.9(4) & 6.1(16) \\
L1489 & H$_2$CN & 1$_{0,1}$-0$_{0,0}$ $J$\,=\,1.5-0.5 $F_1$\,=\,2.5-1.5 $F$\,=\,2.5-1.5 & 73345.486 & 6.49(5) & 0.49(13) & 7.8(2) & 4.1(9) \\
L1489 & H$_2$CN & 1$_{0,1}$-0$_{0,0}$ $J$\,=\,1.5-0.5 $F_1$\,=\,2.5-1.5 $F$\,=\,3.5-2.5 & 73349.648 & 6.86(5) & 0.33(8) & 9.1(2) & 3.2(8) \\
\hline
\end{tabular}
\tablenotea{The line parameters $V_{\rm LSR}$, $\Delta v$ (full width at half maximum), $T_A^*$ peak, and $\int T_A^* dv$ were derived from a Gaussian fit to the line profile. Numbers in parentheses are 1$\sigma$ uncertainties in units of the last digits. $^a$ Marginal detection.\\
}
\end{table*}

The astronomical observations were carried out with the IRAM\,30m telescope during December 2021, April 2022, and June 2022. The weather conditions ranged from very good to moderate, with levels of precipitable water vapor in the range 1-10 mm. We used the EMIR E090 receiver connected to a fast Fourier transform spectrometer providing a spectral resolution of 48.8 kHz. We targeted the 1$_{0,1}$-0$_{0,0}$ rotational transition of ortho H$_2$NC, which consists of multiples hyperfine components around 72.2 GHz, and the 1$_{0,1}$-0$_{0,0}$ transition of ortho H$_2$CN, which has the strongest hyperfine components around 73.3 GHz. For both H$_2$NC and H$_2$CN, ortho levels have $K_a$ even, while para levels have $K_a$ odd.

We searched for H$_2$NC toward the cold dense clouds L1544, L134N, TMC-2, Lupus-1A, L1489, TMC-1 at the peak position of NH$_3$ (hereafter TMC-1\,NH$_3$), L1498, and L1641N, and toward the diffuse clouds B0355+508 and B0415+379. The source coordinates are given in Table~\ref{table:coordinates}. The criteria to select these sources were bright emission of HNC, based on the expected analogy of the H$_2$NC/H$_2$CN and HNC/HCN ratios, and bright emission of NH$_3$, because of the potential correlation between H$_2$NC and NH$_3$. We used the studies of cold dense clouds by \cite{Suzuki1992} for ammonia data and by \cite{Hirota1998} for HNC data. We also included the dense core Lupus-1A because of its similarity to TMC-1\,CP \citep{Sakai2010}. Regarding diffuse clouds, we included the two clouds among the sample of \cite{Liszt2001} with the deepest absorption of HNC, B0415+379 and B0355+508, both of which show also deep absorption by NH$_3$ \citep{Liszt2006}.

For the cold dense clouds the observations were carried out in the frequency-switching observing mode with a frequency throw of 7.2 MHz, while for the diffuse clouds we used the wobbler-switching observing mode with the secondary mirror nutating by 3$'$ at a rate of 0.5 Hz. Focus was checked at the beginning of each observing session and after sunrise and sunset and pointing was regularly checked every 1 h. The intensity scale is expressed in terms of $T_A^*$, the antenna temperature corrected for atmospheric absorption and for antenna ohmic and spillover losses. The uncertainty in $T_A^*$ is estimated to be around 10\,\%. All data were reduced using the program CLASS from the software GILDAS\footnote{https://www.iram.fr/IRAMFR/GILDAS/}.

\begin{figure*}
\centering
\includegraphics[angle=0,width=0.85\textwidth]{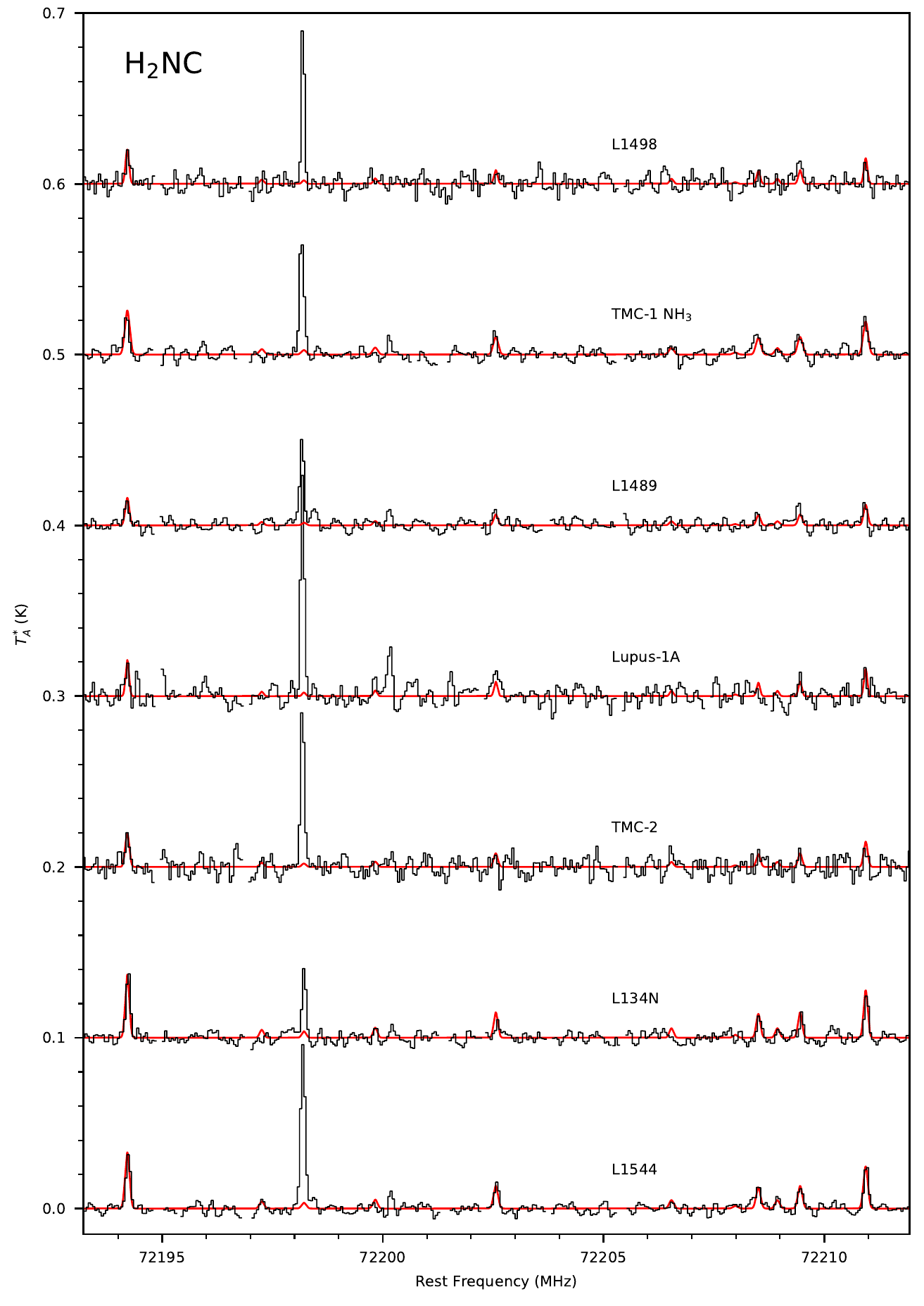}
\caption{Spectra of L1544, L134N, TMC-2, Lupus-1A, L1489, TMC-1\,NH$_3$, and L1498 in the region covering the two most intense hyperfine components of the 1$_{0,1}$-0$_{0,0}$ transition of H$_2$NC, at 72194.211 MHz and 72210.940 MHz. In red we show the calculated synthetic spectra for H$_2$NC adopting the column densities given in Table~\ref{table:column_densities}, an ortho-to-para ratio of 3, a rotational temperature of 4.0 K, a line width consistent with the values derived from the observations (see Table~\ref{table:lines}), and an emission size that fills the IRAM\,30m main beam. The bright line around 72198 MHz corresponds to C$_2$D (see \citealt{Cabezas2021}).} \label{fig:h2nc}
\end{figure*}

\begin{figure}
\centering
\includegraphics[angle=0,width=0.95\columnwidth]{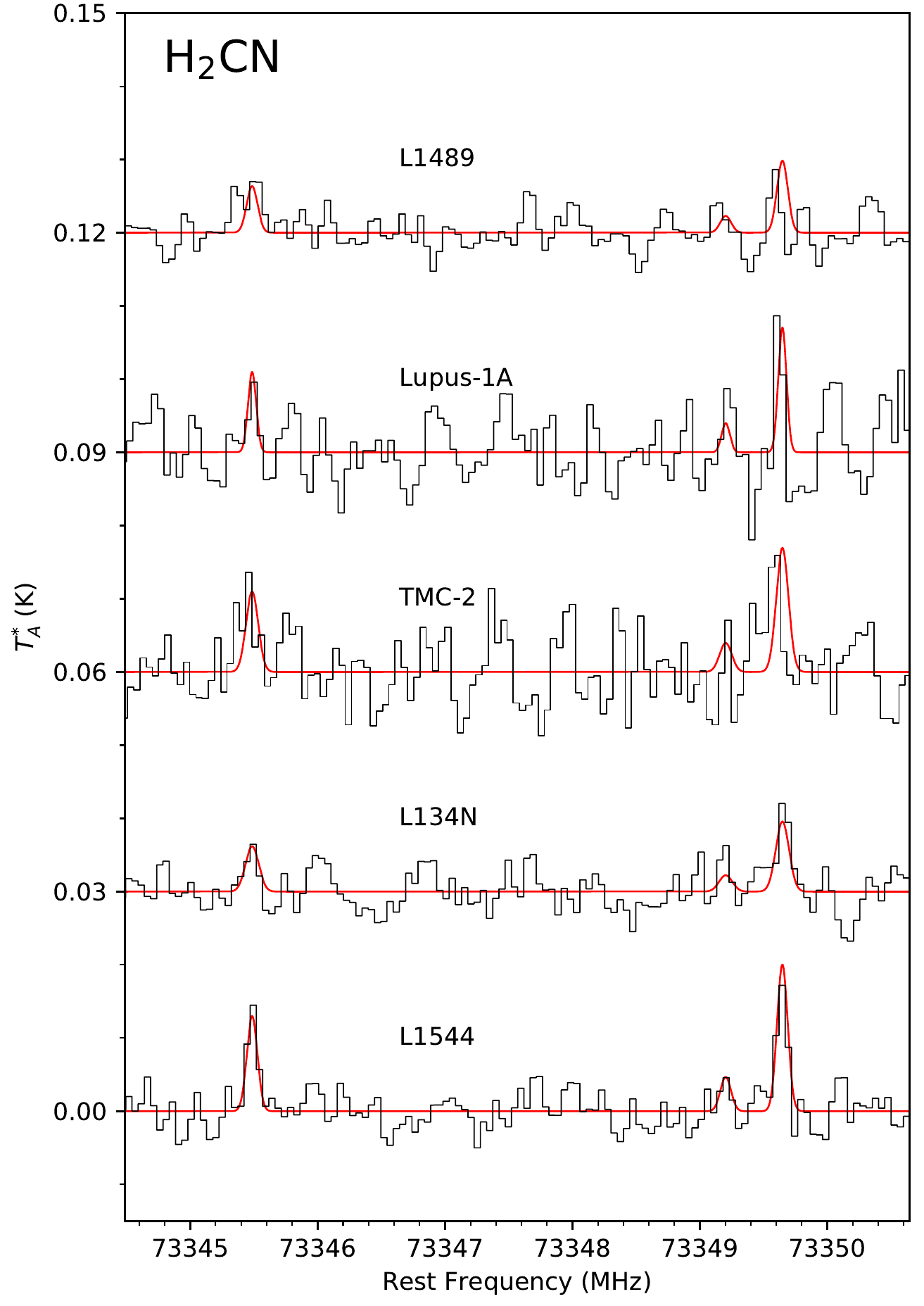}
\caption{Spectra of L1544, L134N, TMC-2, Lupus-1A, and L1489 in the region covering two of the brightest hyperfine components of the 1$_{0,1}$-0$_{0,0}$ transition of H$_2$CN, lying at 73345.486 MHz and 73349.648 MHz. In red we superimpose the computed synthetic spectra for H$_2$CN adopting the column densities given in Table~\ref{table:column_densities}, an ortho-to-para ratio of 3, a rotational temperature of 4.0 K, a line width consistent with the values derived from the observations (see Table~\ref{table:lines}), and an emission size that fills the IRAM\,30m main beam.} \label{fig:h2cn}
\end{figure}

\begin{figure}
\centering
\includegraphics[angle=0,width=0.95\columnwidth]{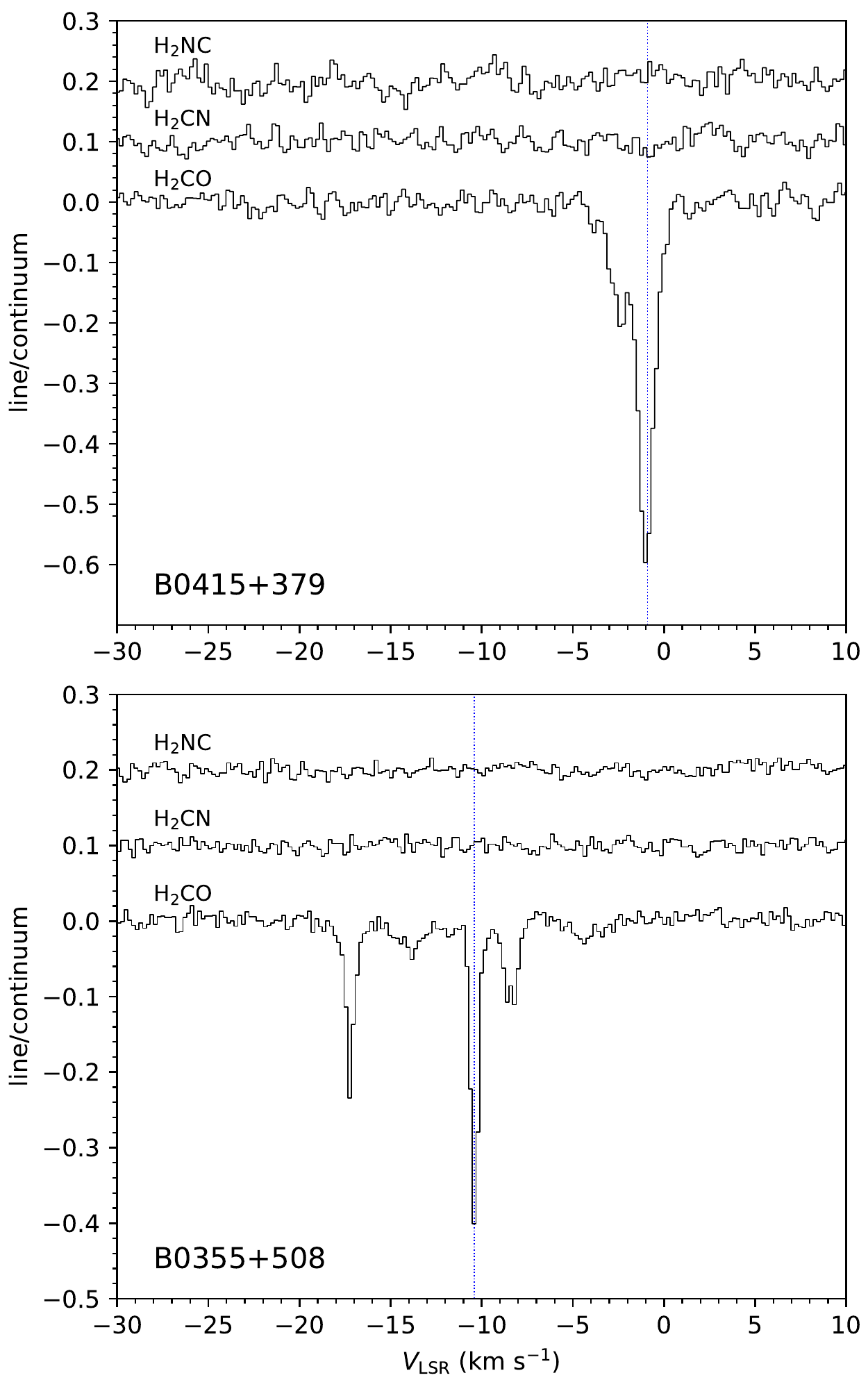}
\caption{Spectra of B0415+379 and B0355+508 at the frequencies of H$_2$CO 1$_{0,1}$-0$_{0,0}$ (72837.949 MHz) and of the strongest hyperfine component of H$_2$NC and H$_2$CN (72194.211 MHz and 73349.648 MHz, respectively). The $y$ axis is expressed as line/continuum. The position of the deepest absorption of H$_2$CO, $V_{\rm LSR}$\,=\,$-$0.9 km s$^{-1}$ in B0415+379 and $V_{\rm LSR}$\,=\,$-$10.4 km s$^{-1}$ in B0355+508, is indicated by a blue dotted vertical line.} \label{fig:diffuse}
\end{figure}

\section{Astronomical results}

\subsection{Detections and non-detections of H$_2$NC and H$_2$CN} \label{sec:detections}

The high-energy isomer H$_2$NC turned out to be easier to detect than its more stable isomer H$_2$CN. We detected H$_2$NC in 7 out of the 8 cold dense clouds targeted (L1544, L134N, TMC-2, Lupus-1A, L1489, TMC-1\,NH$_3$, and L1498; that is, in all but L1641N), while H$_2$CN was only detected in 5 of these sources (L1544, L134N, TMC-2, Lupus-1A, and L1489). On the other hand, neither H$_2$NC nor H$_2$CN were detected in the two diffuse clouds observed, B0415+379 and B0355+508.

We searched for H$_2$NC through the 1$_{0,1}$-0$_{0,0}$ transition, which lies around 72.2 GHz and spreads over $\sim$\,50 MHz due to fine and hyperfine splitting (see \citealt{Cabezas2021}). In Fig.~\ref{fig:h2nc} we show the part of the spectrum covering the two brightest hyperfine components for the 7 cold dense clouds where H$_2$NC was detected. The strongest component at 72194.211 MHz is well detected in the 7 sources, while the second strongest one at 72210.940 MHz is well seen in all sources except TMC-2, where it is only marginally detected. In addition, there are various weaker hyperfine components for which there is a good match between observed and calculated spectra. The line parameters for the two strongest components of H$_2$NC 1$_{0,1}$-0$_{0,0}$ are given in Table~\ref{table:lines}. In the case of H$_2$CN, we targeted the 1$_{0,1}$-0$_{0,0}$ transition, which lies around 73.3 GHz and has also a complex hyperfine pattern with components spread over almost 200 MHz. We show the observed spectra around the most intense component, lying at 73349.648 MHz, for the 5 cold dense clouds where H$_2$CN was detected in Fig.~\ref{fig:h2cn}. In some of the sources the weaker component at 73345.486 MHz is also clearly seen. The line parameters for these two components are given in Table~\ref{table:lines}.

The two isomers H$_2$NC and H$_2$CN were also searched for toward the diffuse clouds B0415+379 and B0355+508, although none of them were detected in these two sources. In Fig.~\ref{fig:diffuse} we show the observed spectra around the strongest hyperfine component of H$_2$NC and H$_2$CN, at 72194.211 MHz and 73349.648 MHz, respectively, for the two clouds. In addition we show the observed spectra around the 1$_{0,1}$-0$_{0,0}$ transition of H$_2$CO, lying at 72837.949 MHz, which shows absorption at different velocities, in agreement with previous observations of other transitions of H$_2$CO toward these two sources \citep{Liszt1995}. The ordinate scale is expressed as line/continuum, following the convention by \cite{Liszt2006}, where by 'line' we mean the $T_A^*$ of the observed spectrum after baseline (or continuum) subtraction and by 'continuum' we mean the $T_A^*$ of the continuum level. We find that the deepest H$_2$CO absorption in B0415+379 occurs at $V_{\rm LSR}$\,=\,$-$0.9 km s$^{-1}$, while in B0355+508 it is located at a velocity of $V_{\rm LSR}$\,=\,$-$10.4 km s$^{-1}$, in agreement with previous observations of simple molecules toward these two sources (e.g., \citealt{Liszt1995,Liszt2001,Liszt2006}). As shown in Fig.~\ref{fig:diffuse}, there is not hint of absorption by H$_2$NC or H$_2$CN toward either B0415+379 or B0355+508.

\subsection{Calculation of column densities}

We calculated column densities for H$_2$NC and H$_2$CN in those sources where the molecule was detected. When the molecule was not detected, we computed an upper limit to the column density.

\begin{table*}
\small
\caption{Column densities of H$_2$NC and H$_2$CN.}
\label{table:column_densities}
\centering
\begin{tabular}{l@{\hspace{1.5cm}}ccccc@{\hspace{1.5cm}}cc}
\hline \hline
\multicolumn{1}{l}{Source} & \multicolumn{1}{c}{$N$(H$_2$NC)} & \multicolumn{1}{c}{$N$(H$_2$CN)} & \multicolumn{1}{c}{H$_2$NC/H$_2$CN} & \multicolumn{1}{c}{Reference} & & \multicolumn{1}{c}{$N$(NH$_3$)} & \multicolumn{1}{c}{Reference} \\
& \multicolumn{1}{c}{(cm$^{-2}$)} & \multicolumn{1}{c}{(cm$^{-2}$)} & & & &  \multicolumn{1}{c}{(cm$^{-2}$)} & \\
\hline
\multicolumn{8}{c}{cold dense clouds} \\
\hline
L483            &      1.0\,$\times$\,10$^{12}$ &      8.0\,$\times$\,10$^{11}$ &     1.25 & (1) & &     14\,$\times$\,10$^{14}$ & (3) \\
B1-b            &      1.0\,$\times$\,10$^{12}$ &      1.0\,$\times$\,10$^{12}$ &     1.00 & (1) & &   11.2\,$\times$\,10$^{14}$ & (4) \\
TMC-1\,CP       & $<$\,3.2\,$\times$\,10$^{11}$ & $<$\,4.8\,$\times$\,10$^{11}$ &       -- & (1) & &    1.9\,$\times$\,10$^{14}$ & (4) \\
L1544           &      3.9\,$\times$\,10$^{11}$ &      4.3\,$\times$\,10$^{11}$ &     0.91 & (2) & &    2.7\,$\times$\,10$^{14}$ & (4) \\
L134N           &      4.0\,$\times$\,10$^{11}$ &      2.5\,$\times$\,10$^{11}$ &     1.60 & (2) & &    7.8\,$\times$\,10$^{14}$ & (4) \\
TMC-2           &      2.1\,$\times$\,10$^{11}$ &      4.1\,$\times$\,10$^{11}$ &     0.51 & (2) & &    6.1\,$\times$\,10$^{14}$ & (4) \\
Lupus-1A        &      1.9\,$\times$\,10$^{11}$ &      2.8\,$\times$\,10$^{11}$ &     0.68 & (2) & &    1.1\,$\times$\,10$^{14}$ & (5) \\
L1489           &      1.7\,$\times$\,10$^{11}$ &      2.1\,$\times$\,10$^{11}$ &     0.81 & (2) & &    6.3\,$\times$\,10$^{14}$ & (4) \\
TMC-1\,NH$_3$   &      3.3\,$\times$\,10$^{11}$ & $<$\,1.2\,$\times$\,10$^{11}$ & $>$\,2.7 & (2) & &    5.2\,$\times$\,10$^{14}$ & (4) \\
L1498           &      1.9\,$\times$\,10$^{11}$ & $<$\,1.4\,$\times$\,10$^{11}$ & $>$\,1.4 & (2) & &    4.1\,$\times$\,10$^{14}$ & (4) \\
L1641N          & $<$\,3.1\,$\times$\,10$^{11}$ & $<$\,4.6\,$\times$\,10$^{11}$ &       -- & (2) & &   11.0\,$\times$\,10$^{14}$ & (4) \\
\hline
\multicolumn{8}{c}{diffuse-like clouds} \\
\hline
PKS\,1830$-$211 &      5.3\,$\times$\,10$^{11}$ &      2.0\,$\times$\,10$^{12}$ &     0.27 & (1) & & (5-10)\,$\times$\,10$^{14}$ & (6) \\
B0415+379       & $<$\,3.1\,$\times$\,10$^{11}$ & $<$\,5.5\,$\times$\,10$^{11}$ &       -- & (2) & &  10.53\,$\times$\,10$^{12}$ & (7) \\
B0355+508       & $<$\,9.2\,$\times$\,10$^{10}$ & $<$\,2.0\,$\times$\,10$^{11}$ &       -- & (2) & &   2.54\,$\times$\,10$^{12}$ & (7) \\
\hline
\end{tabular}
\tablenoteb{References: (1) \cite{Cabezas2021}. (2) This work. (3) \cite{Anglada1997}. (4) \cite{Suzuki1992}. (5) Value at position Lup1\,C4, $\sim$\,2\,arcmin off from Lupus-1A \citep{Benedettini2012}. (6) \cite{Henkel2008}. (7) Values at the velocity components $V_{\rm LSR}$\,=\,$-$0.9 km s$^{-1}$ for B0415+379 and $V_{\rm LSR}$\,=\,$-$10.4 km s$^{-1}$ for B0355+508 \citep{Liszt2006}.}
\end{table*}

For the cold dense clouds L1544, L134N, TMC-2, Lupus-1A, L1489, TMC-1\,NH$_3$, L1498, and L1641N we computed H$_2$NC and H$_2$CN column densities assuming local thermodynamic equilibrium (LTE). Since the hyperfine components of H$_2$NC and H$_2$CN observed belong to the same rotational transition, the 1$_{0,1}$-0$_{0,0}$ for both molecules, the detected lines have very similar upper level energies and thus it is not possible to determine the rotational temperature. In L483 \cite{Cabezas2021} derive a rotational temperature of 4.0 K for both H$_2$NC and H$_2$CN from the observation of various rotational transitions with different upper level energies.This value is lower than the gas kinetic temperature (around 10 K in L483; \citealt{Anglada1997,Agundez2019}), which indicates subthermal excitation, in line with the fact that these two molecules have high dipole moments (3.83 D for H$_2$NC and 2.54 D for H$_2$CN; \citealt{Cabezas2021,Cowles1991}). Since cold dense clouds typically have kinetic temperatures around 10 K and H$_2$ volume densities of a few 10$^4$ cm$^{-3}$ \citep{Benson1989}, in the absence of better constraints we adopted the rotational temperature derived in L483 for both H$_2$NC and H$_2$CN, 4.0 K, in all the cold dense clouds. We however note that the excitation of H$_2$NC and H$_2$CN could be different, as found for other isomer such as HCN and HNC \citep{HernandezVera2017} or the family of isomers of cyanoacetylene \citep{Bop2019}. In the case of H$_2$NC and H$_2$CN, collision rate coefficients are not available and thus we cannot investigate this point, although we acknowledge that this introduces an uncertainty in the calculation of the column densities. \cite{Cabezas2021} also showed that the ortho-to-para ratio of H$_2$NC in L483 is consistent with the statistical value of three. We adopt this value here for both H$_2$NC and H$_2$CN, although we note that this is an additional source of uncertainty in the calculation of the column densities. With these assumptions, we computed synthetic spectra for H$_2$NC and H$_2$CN adopting as line width a value consistent with that observed in each source (see Table~\ref{table:lines}). The synthetic spectra calculated for H$_2$NC and H$_2$CN are compared with the observed ones in Fig.~\ref{fig:h2nc} and Fig.~\ref{fig:h2cn}, respectively, while the corresponding column densities derived are given in Table~\ref{table:column_densities}.

In the case of non-detections of either H$_2$NC or H$_2$CN in cold dense clouds we derived 3\,$\sigma$ upper limits to the column density, which are also given in Table~\ref{table:column_densities}. The only reported detection of H$_2$NC or H$_2$CN in the sources observed here is that of H$_2$CN in L1544 by \cite{Vastel2019}. The column density derived here, 3.9\,$\times$\,10$^{11}$ cm$^{-2}$ (see Table~\ref{table:column_densities}), is in agreement with the range of values found by \cite{Vastel2019}, (3-6)\,$\times$\,10$^{11}$ cm$^{-2}$.

In the case of the diffuse clouds B0415+379 and B0355+508 we derived upper limits to the column densities of H$_2$NC and H$_2$CN in the following way. We first computed the line opacity $\tau$ using the equation (e.g., \citealt{Linke1981,Nyman1984,Greaves1992,Corby2018})
\begin{equation}
\tau = \rm - Ln \bigg( 1 + \frac{line}{continuum} \bigg). \label{eq:tau}
\end{equation}
The velocity-integrated opacity $\int \tau dv$ is then related to the column density $N$ through the expression \citep{Nyman1984,Greaves1996,Corby2018}
\begin{equation}
\int \tau dv = \frac{N}{Q(T_{rot})} \frac{8 \pi^3 S \mu^2}{3 h} \bigg( e^{-E_l/k T_{rot}} - e^{-E_u/k T_{rot}} \bigg), \label{eq:n}
\end{equation}
where $T_{rot}$ is the rotational temperature, $Q(T_{rot})$ is the partition function, $S$ is the line strength, $\mu$ is the dipole moment, $E_l$ and $E_u$ are the energies of the lower and upper level, respectively, and $h$ and $k$ are the Planck and Boltzmann constants, respectively. For convenience, Eq.~(\ref{eq:n}) can be expressed as
\begin{equation}
\int \tau dv = \frac{N}{Q(T_{rot})} \frac{S \mu^2}{8.0 \times 10^{12}} \bigg( e^{-E_l/T_{rot}} - e^{-E_u/T_{rot}} \bigg), \label{eq:n_units}
\end{equation}
where $\int \tau dv$ is in units of km s$^{-1}$, $N$ in cm$^{-2}$, $\mu$ in Debye, and $E_l$, $E_u$, and $T_{rot}$ are in units of K. We assumed a rotational temperature equal to that of the cosmic microwave background, i.e., 2.725 K. To derive upper limits to the column densities of H$_2$NC and H$_2$CN we used the line parameters of the strongest hyperfine component, lying at 72194.211 MHz and 73349.648 MHz, respectively. For H$_2$NC, $\mu$\,=\,3.83 D, $Q(2.725 K)$\,=\,35.075, $S$\,=\,2.67, $E_l$\,=\,0.012 K, and $E_u$\,=\,3.477 K, while for H$_2$CN, $\mu$\,=\,2.54 D, $Q(2.725 K)$\,=\,34.597, $S$\,=\,2.67, $E_l$\,=\,0.018 K, and $E_u$\,=\,3.538 K. The 3\,$\sigma$ upper limits to the column densities of H$_2$NC and H$_2$CN in B0415+379 and B0355+508 are given in Table~\ref{table:column_densities}. As a reference, the column densities derived for para-H$_2$CO are 4.0\,$\times$\,10$^{12}$ cm$^{-2}$ in B0415+379 at $V_{\rm LSR}$\,=\,$-$0.9 km s$^{-1}$ and 1.1\,$\times$\,10$^{12}$ cm$^{-2}$ in B0355+508 at $V_{\rm LSR}$\,=\,$-$10.4 km s$^{-1}$, values which are very similar to those derived by \cite{Liszt1995}.

\section{Astrophysical lessons on H$_2$NC and H$_2$CN} \label{sec:discussion}

The column densities derived for H$_2$NC and H$_2$CN are typically in the range 10$^{11}$-10$^{12}$ cm$^{-2}$, where the largest values are found for the cold dense clouds L483 and B1-b and the $z$\,=\,0.89 galaxy in front of PKS\,1830$-$211, which happen to be the three sources where H$_2$NC was initially discovered by \cite{Cabezas2021}. On the other hand, the H$_2$NC/H$_2$CN abundance ratio is relatively uniform around unity, with the lowest value of 0.27 found toward PKS\,1830$-$211 and the largest one, $>$\,2.7, occurring in the cold dense cloud TMC-1\,NH$_3$. We aim to understand whether the abundances of H$_2$NC and H$_2$CN, or the abundance ratio H$_2$NC/H$_2$CN, depend on some physical or chemical characteristic of the host cloud.

\begin{figure*}
\centering
\includegraphics[angle=0,width=0.78\textwidth]{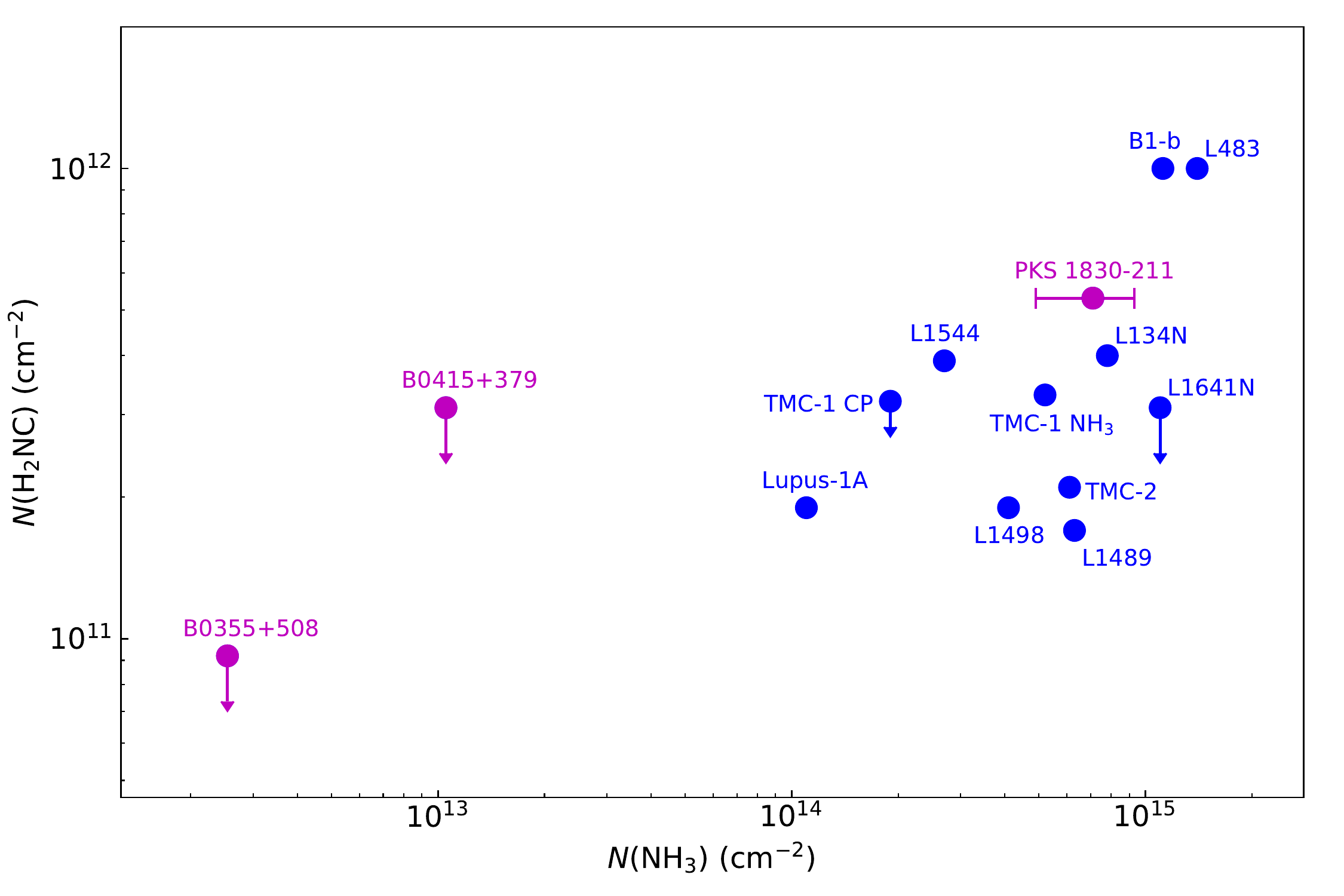}
\caption{The column density of H$_2$NC is represented as a function of the column density of NH$_3$ for the sources where H$_2$NC has been searched for. Cold dense clouds are plotted in blue and diffuse-like clouds in magenta. The values of the column densities and the corresponding references are given in Table~\ref{table:column_densities}.} \label{fig:correlation}
\end{figure*}

\cite{Cabezas2021} proposed that the H$_2$NC/H$_2$CN ratio could behave as the HNC/HCN ratio, with values close to one in cold dense clouds and below one in diffuse clouds. This idea was suggested by the fact that the H$_2$NC/H$_2$CN ratio derived in the cold dense clouds L483 and B1-b was $\sim$\,1, while in the redshifted galaxy in front of the quasar PKS\,1830$-$211, which has a chemical composition characteristic of a diffuse or translucent cloud, the H$_2$NC/H$_2$CN ratio was found to be 0.27. With this idea in mind, in this study we searched for H$_2$NC and H$_2$CN in various cold dense clouds and in the two most promising diffuse clouds known to contain a relatively rich molecular content, based on the series of studies by Liszt and Lucas (e.g., \citealt{Liszt2001,Liszt2006}). The observations of cold dense clouds carried out here confirm that the H$_2$NC/H$_2$CN ratio is close to one in this type of sources, with the largest deviation occurring for TMC-1\,NH$_3$, where the detection of H$_2$NC and the non detection of H$_2$CN set a lower limit to the H$_2$NC/H$_2$CN of 2.7. However, we did not detect H$_2$NC or H$_2$CN toward the diffuse clouds B0415+379 and B0355+508, making impossible to constrain the H$_2$NC/H$_2$CN ratio in galactic diffuse clouds. The non detection of H$_2$NC and H$_2$CN in these two diffuse clouds is probably related to a too low column density of NH$_3$, in comparison with the typical column densities of NH$_3$ in cold dense clouds, which are 1-2 orders of magnitude higher (see discussion on this point below). The sensitivity achieved in the IRAM\,30m observations of these two diffuse clouds, in particular for B0415+379 where the $T_A^*$ rms noise level reached is $\sim$\,4 mK, suggests that it would be challenging to detect either H$_2$NC or H$_2$CN in a galactic diffuse cloud. It seems therefore difficult to confirm whether or not the H$_2$NC/H$_2$CN ratio is significantly lower in diffuse clouds than in cold dense clouds.

Another interesting feature discussed by \cite{Cabezas2021} concerns the chemical origin of H$_2$NC. The reactions N + CH$_3$ and C + NH$_3$ are two potentially important sources of H$_2$NC. Chemical routes to H$_2$NC involving ions are also possible. For example, the dissociative recombination of cations, such as CH$_3$N$^+$, CH$_2$NH$_2^+$, CH$_3$NH$_2^+$, or CH$_3$NH$_3^+$, with electrons could produce as fragments H$_2$NC and H$_2$CN, although experimental or theoretical information on these reactions is scarce \citep{Yuen2019}. A similar mechanism operates in cold dense clouds for HCN and HNC, with the two isomers being formed with nearly equal yields upon dissociative recombination of HCNH$^+$ with electrons \citep{Mendes2012}. The reaction C + NH$_3$ is a more likely source of H$_2$NC than N + CH$_3$ because it involves a smaller rearrangement of atoms. If the reaction C + NH$_3$ is the main pathway to H$_2$NC we could expect a relation between the abundance of H$_2$NC and that of NH$_3$. To investigate this point we represented in Fig.~\ref{fig:correlation} the column density of H$_2$NC versus that of NH$_3$. Some interesting features emerge from this plot. First, the two sources with the largest column densities of H$_2$NC, which are L483 and B1-b, have also the highest column densities of NH$_3$. Second, among the cold dense clouds, the sources with the lowest column densities of NH$_3$, Lupus-1A and TMC-1\,CP, have also low column densities of H$_2$NC. These two findings are very suggestive of a relation between H$_2$NC and NH$_3$, and thus of NH$_3$ being a precursor of H$_2$NC. Cold dense clouds with NH$_3$ column densities in between those of Lupus-1A and TMC-1\,CP, on one side, and L483 and B1-b, on the other, show a less clear relation between H$_2$NC and NH$_3$. In particular, there are sources, such as L1498, L1489, TMC-2, and L1641N, which have a moderately high column density of NH$_3$ but a low H$_2$NC column density. This fact suggests that NH$_3$ may not be the only species that regulates the abundance of H$_2$NC. Indeed, if H$_2$NC is mainly formed by the reaction C + NH$_3$, the abundance of atomic carbon would also be a relevant parameter setting the abundance of H$_2$NC. However, it is very difficult to constrain the abundance of atomic carbon inside cold dense clouds \citep{Schilke1995,Stark1996}. Diffuse clouds have a markedly different chemistry compared to cold dense clouds, with a lower chemical complexity (e.g., \citealt{Liszt2008}). This may be the reason why H$_2$NC and H$_2$CN were not detected toward B0415+379 or B0355+508. However, if the relation between H$_2$NC and NH$_3$ found for cold dense clouds holds down to the diffuse stage, the reason of the non detection of H$_2$NC and H$_2$CN toward B0415+379 or B0355+508 could be the low NH$_3$ column density in these two sources, 1-2 orders of magnitude smaller than in cold dense clouds (see Fig.~\ref{fig:correlation}).

\section{The reaction C + NH$_3$ as a source of H$_2$NC}

The suspected chemical link between H$_2$NC and NH$_3$ in interstellar clouds may be attributed to the fact that the reaction C + NH$_3$ is the main formation mechanism of H$_2$NC in these environments. Here we revisit this reaction and investigate whether H$_2$NC can be produced this way.

\subsection{Previous studies}

\begin{figure*}
\centering
\includegraphics[angle=0,width=0.80\textwidth]{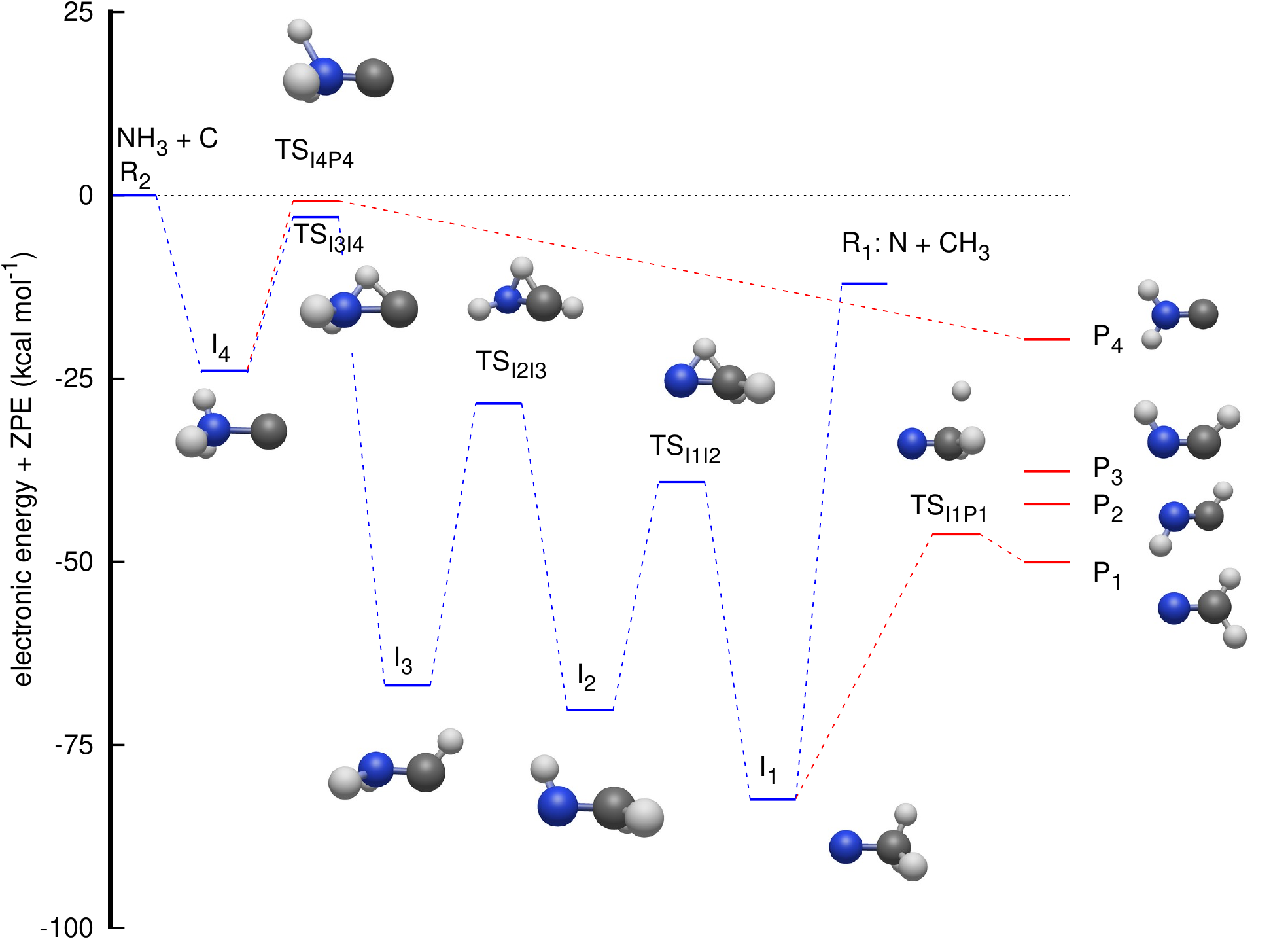}
\caption{Energies (in kcal mol$^{-1}$) of the stationary points describing the C + NH$_3$ reaction on the triplet state obtained at RCCSD(T)-F12 with the VQZ-F12 basis set, including the ZPE. All energies are referred to the C($^3P$) + NH$_3$($^1A'$) reactants.} \label{fig:pes}
\end{figure*}

The reaction C + NH$_3$ has been found to be fast at low temperature, with a measured rate coefficient of (1.8\,$\pm$\,0.2)\,$\times$\,10$^{-10}$ cm$^3$ s$^{-1}$ at 50 K \citep{Hickson2015}. The reaction products,
\begin{subequations} \label{reac:c+nh3}
\begin{align} 
\rm C(^3P) + NH_3(^1A') & \rightarrow \rm H(^2S) + H_2CN(^2B_2) \label{reac:c+nh3:h2cn} \\
                                         & \rightarrow \rm H(^2S) + \textit{trans-}HCNH(^2A') \\
                                         & \rightarrow \rm H(^2S) + \textit{cis-}HCNH(^2A') \\
                                         & \rightarrow \rm H(^2S) + H_2NC(^2B_2),
\end{align}
\end{subequations}
were investigated by \cite{Bourgalais2015}. These authors measured the H atom yield to be 0.99\,$\pm$\,0.08, independent of temperature, over the 50-296 K range. On the other hand, they also used tunable vacuum-ultraviolet photoionization coupled with time of flight mass spectrometry in a gas flow tube experiment at 330 K, which indicated that H$_2$CN, and not \textit{cis} or \textit{trans} HCNH, is likely the main isomer produced. \cite{Bourgalais2015} also performed electronic structure calculations, which indicated that the reactants approach each other in an attractive potential without barrier to form the intermediate CNH$_3$ (hereafter I$_4$; see Fig.~\ref{fig:pes}). From I$_4$ the dynamics can proceed overpassing two barriers. The lower is TS$_{\rm I3I4}$, that connects I$_4$ to I$_3$, giving access to the statistical reaction mechanism that is expected to majoritarily lead to H$_2$CN. The second barrier, TS$_{\rm I4P4}$, connects I$_4$ directly with the H$_2$NC product, and according to \cite{Bourgalais2015} has an energy slightly above the reactants (although this is very method dependent), which implies that this pathway is closed at a temperature of 10 K. As a consequence, \cite{Bourgalais2015} assigned a branching ratio of 100\,\% for channel~(\ref{reac:c+nh3:h2cn}).

It is also interesting to discuss the reaction N + CH$_3$ because it occurs on the same triplet potential energy surface (PES) than C + NH$_3$. This reaction has been studied experimentally \citep{Stief1988,Marston1989a,Marston1989b} and theoretically \citep{Gonzalez1992,Sadygov1997,Cimas2006,Alves2008,Chiba2015}. The rate coefficient has been measured to be (0.64-1.7)\,$\times$\,10$^{-10}$ cm$^3$ s$^{-1}$ in the temperature range 200-423 K \citep{Marston1989a}. The major reaction products have been experimentally inferred to be H$_2$CN + H, with about 10\,\% of the reaction leading to HCN + H$_2$ \citep{Marston1989b}, which has been theoretically explained in terms of an intersystem crossing from the triplet to the singlet PES \citep{Gonzalez1992,Sadygov1997,Chiba2015}. The mechanism of the reaction has been discussed in detail by \cite{Cimas2006}. The reactants approach each other in an attractive potential to form the intermediate NCH$_3$ (hereafter I$_1$; see Fig.~\ref{fig:pes}), which is connected to two saddle points: the lower is TS$_{\rm I1P1}$, which yields H$_2$CN as product, while the higher TS$_{\rm I1I2}$ connects to the HCNH$_2$ intermediate (I$_2$). The two barriers are submerged, i.e., have energies below the reactants N + CH$_3$, but since TS$_{\rm I1P1}$ $<$ TS$_{\rm I1I2}$, according to RRKM arguments most of the flux will go through TS$_{\rm I1P1}$ to form H$_2$CN. The I$_2$ intermediate in turn is connected to I$_3$ (H$_2$NCH). The path along these intermediates I$_1$ $\rightarrow$ I$_2$ $\rightarrow$ I$_3$, in which successive H atom migrations take place, is expected to behave statistically, and therefore the main product along this second statistical reaction mechanism is also expected to be H$_2$CN. Thus, \cite{Cimas2006} conclude that the main products should be H + H$_2$CN.

\subsection{Stationary points of the triplet electronic state}

We carried out a progressive optimization of the stationary points of the triplet PES using several methods, M06-2X, RCCSD(T), and RCCSD(T)-F12, and different basis sets, VTZ-F12 and VQZ-F12 \citep{Peterson2008}, using the MOLPRO package \citep{Werner2012}. The relative energies of the stationary points varied considerably depending on the method and basis set, specially for the transition states, which indicates that electronic correlation is very important. For every transition state we calculated the intrinsic reaction coordinate to check the two minima connected to it.

\begin{table}
\small
\caption{Electronic energies $E_{el}$ calculated at RCCSD(T)-F12/VQZ-F12 level, zero point energies (ZPE), and the sum of both for the stationary points involved in the C + NH$_3$ reaction on the triplet state. Energies are referred to C + NH$_3$ (R$_2$). All values in kcal mol$^{-1}$.}
\label{table:pes}
\centering
\begin{tabular}{ll@{\hspace{0.62cm}}c@{\hspace{0.62cm}}c@{\hspace{0.62cm}}c}
\hline \hline
Name & System & $E_{el}$ & ZPE & $E_{el}$ + ZPE \\
\hline
R$_2$           & C($^3$P)+NH$_3$($^1A'$)    &     0.0   & 21.599 &     0.0   \\
R$_1$           & N($^4$S)+CH$_3$($^2A_2''$) &  $-$9.153 & 18.718 & $-$12.033 \\
\hline
I$_4$           & C-NH$_3$                   & $-$26.884 & 24.580 & $-$23.902 \\
I$_3$           & HC-NH$_2$                  & $-$68.771 & 23.487 & $-$66.882 \\
I$_2$           & H$_2$C-NH                  & $-$71.036 & 22.392 & $-$70.235 \\
I$_1$           & H$_3$C-N                   & $-$84.210 & 23.353 & $-$82.455 \\
\hline
P$_4$           & H + H$_2$NC                & $-$14.745 & 16.664 & $-$19.680 \\
P$_3$           & H + \textit{cis-}HCNH      & $-$31.780 & 15.651 & $-$37.727 \\
P$_2$           & H + \textit{trans-}HCNH    & $-$36.775 & 16.244 & $-$42.129 \\
P$_1$           & H + H$_2$CN                & $-$44.274 & 15.780 & $-$50.092 \\
\hline
TS$_{\rm I3I4}$ & CHNH$_2$                   &  $-$1.750 & 20.420 &  $-$2.929 \\
TS$_{\rm I2I3}$ & HCHNH                      & $-$26.836 & 20.007 & $-$28.427 \\
TS$_{\rm I1I2}$ & H$_2$CHN                   & $-$37.337 & 19.825 & $-$39.111 \\
TS$_{\rm I4P4}$ & CNH$_2$-H                  &     1.547 & 19.306 &  $-$0.745 \\
TS$_{\rm I1P1}$ & H-H$_2$CN                  & $-$41.726 & 17.062 & $-$46.262 \\
\hline
\end{tabular}
\tablenotec{The calculation of $E_{el}$ for N + CH$_3$ (R$_1$) was done independently for each subsystem and their energies were added. The imaginary frequencies at the saddle points TS$_{\rm I3I4}$, TS$_{\rm I2I3}$, TS$_{\rm I1I2}$, TS$_{\rm I4P4}$, TS$_{\rm I1P1}$ are 4.837, 5.719, 5.578, 5.154, and 2.283 kcal mol$^{-1}$, respectively.}
\end{table}

The RCCSD(T)-F12/VQZ-F12 results are listed in Table~\ref{table:pes} and represented in Fig.~\ref{fig:pes}. The height of TS$_{\rm I4P4}$ is critical since it connects the CNH$_3$ well, I$_4$, with the H$_2$NC product, P$_4$. With the RCCSD(T)-F12/VQZ-F12 method the electronic energy of TS$_{\rm I4P4}$ is above that of the reactants C + NH$_3$, but when the zero point energy (ZPE) is included the energy at the TS$_{\rm I4P4}$ point becomes negative, and therefore the H$_2$NC product becomes accessible in the C + NH$_3$ reaction even at low temperatures.

The computational effort of geometry optimization increases dramatically with the size of the basis set, specially when the gradients and hessians need to be calculated numerically, as it happens when using RCCSD(T)-F12 methods. An alternative is to extrapolate the energies obtained with two or more electronic correlation-consistent basis sets to the complete basis set (CBS) limit. There are many extrapolation methods, and some of them were reviewed and applied to H$_3^+$ by \cite{Velilla2010}, who found that the best option is the two points extrapolation method \citep{Lee2000,Huh2003}:
\begin{equation}
E_X = E_{CBS} + A (X+k)^{-3}, \label{eq:cbs}
\end{equation}
where $X$ refers to the basis ($X$\,=\,3 and 4 for VTZ and VQZ, respectively) and $k$\,=\,$-$1 is the best choice when using RCCSD(T) methods. The results are listed in Table~\ref{table:cbs}. In the CBS limit, the transition states TS$_{\rm I4P4}$ and TS$_{\rm I3I4}$ have similar energies, both lying below the reactants C + NH$_3$, as found with the RCCSD(T)-F12/VQZ-F12 method.

\begin{table}
\small
\caption{Electronic energies calculated at the complete basis set (CBS) limit $E_{CBS}$ using the RCCSD(T)-F12, MRCI-F12, and MRCI-F12+Q methods plus the ZPE. Energies are referred to C + NH$_3$ (R$_2$). All values in kcal mol$^{-1}$}
\label{table:cbs}
\centering
\begin{tabular}{l@{\hspace{0.78cm}}c@{\hspace{0.78cm}}c@{\hspace{0.78cm}}c}
\hline \hline
Name & \multicolumn{3}{c}{$E_{CBS}$ + ZPE} \\
\cline{2-4}
 & RCCSD(T)-F12 & MRCI-F12 & MRCI-F12+Q \\
\hline
R$_2$           &     0.0   &     0.0   &     0.0   \\ 
I$_4$           & $-$25.395 & $-$29.574 & $-$28.455 \\
TS$_{\rm I3I4}$ &  $-$3.907 &  $-$7.567 &  $-$7.392 \\
TS$_{\rm I4P4}$ &   $-$1.568 &  $-$7.260 &  $-$6.266 \\
\hline
\end{tabular}
\tablenotec{CBS energies obtained using Eq.~(\ref{eq:cbs}) with the VTZ-F12 and VQZ-F12 basis sets, where in both cases electronic energies are obtained at the geometries optimized using RCCSD(T)-F12/VQZ-F12.}
\end{table}

Single reference methods like RCCSD(T)-F12 have problems to describe reactions, specially at the saddle points, which usually present a large multi-reference character. In the RCCSD(T)-F12 calculations discussed above, the T1-diagnostic is always below 0.023, as previously found by \cite{Bourgalais2015}. This value usually allows to consider that RCCSD(T)-F12 results are reasonably good even at saddle points \citep{Lee1989}. Multi reference methods, such as the multi reference configuration interaction (MRCI) method \citep{Werner1988,Shiozaki2013} are considered more accurate to describe reactions, specially at saddle points. Since MRCI calculations are more computationally demanding, here we performed single point calculations at the geometries optimized with the RCCSD(T)-F12/VQZ-F12 method, with and without the Davidson correction \citep{Davidson1975}, which we denote as MRCI-F12 and MRCI-F12+Q, respectively. The results obtained with the VTZ and VQZ basis sets are then used to get the complete basis limit, which are listed in Table~\ref{table:cbs}. With the MRCI-F12 and MRCI-F12+Q methods the heights of the reaction barriers TS$_{\rm I3I4}$ and TS$_{\rm I4P4}$ are lower than with the RCCSD(T)-F12 method and, interestingly, the energies of these two transitions states become much more similar.

\subsection{Reaction rate coefficient and RKKM branching ratios}

\begin{figure}
\centering
\includegraphics[angle=0,width=\columnwidth]{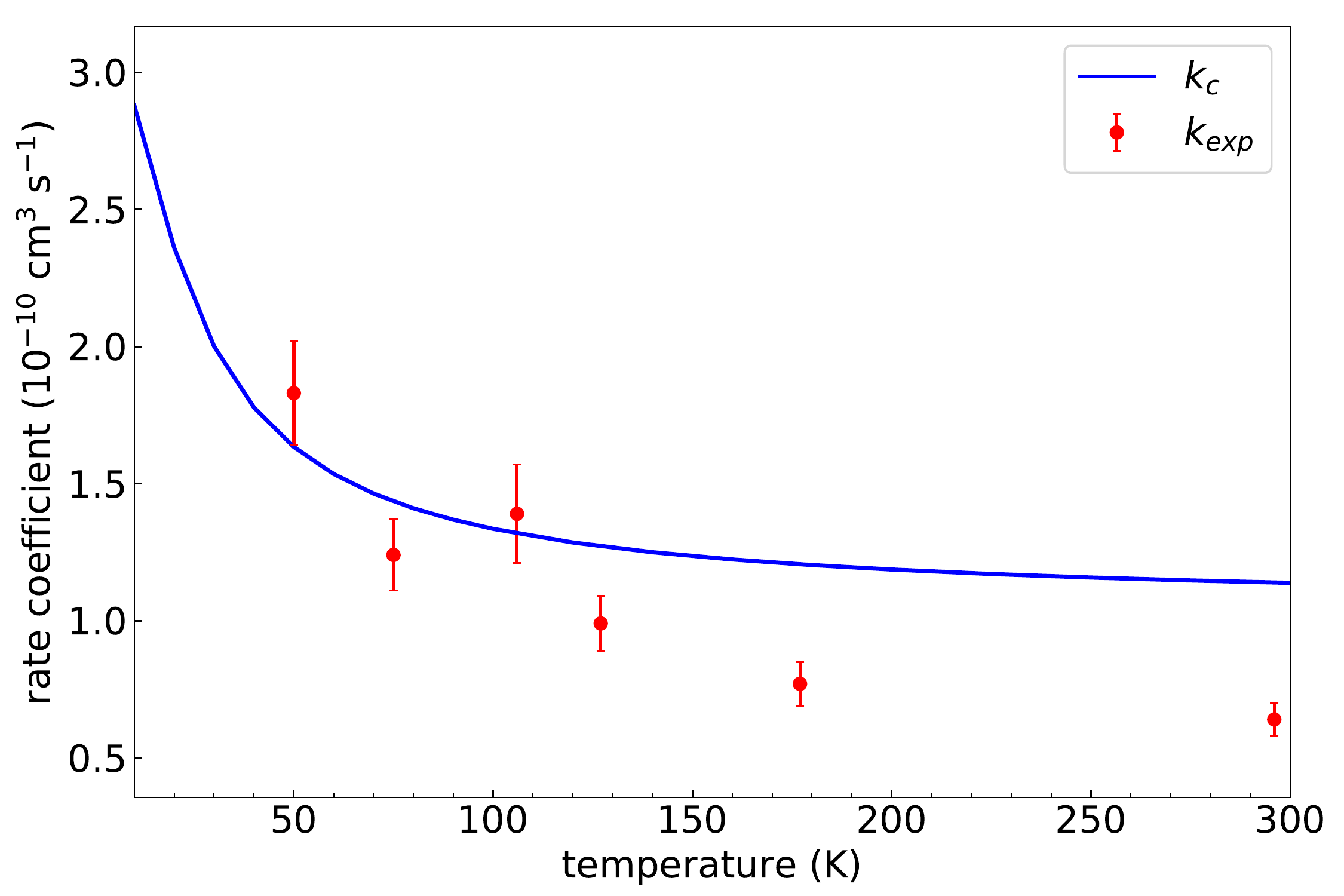}
\caption{Capture rate coefficient $k_c$ calculated for the reaction C + NH$_3$ according to Eq.~(\ref{eq:capture_rate}) is compared with the experimental measurements of the reactive rate coefficient by \cite{Hickson2015}.} \label{fig:rate}
\end{figure}

At low energies, the two reactants approach each other forming the CNH$_3$ intermediate I$_4$ following a descendent potential energy path with no barrier. This is a clear example of a capture mechanism, which is governed by long-range interactions at low temperatures. Long-range forces are due to the interaction of the electric dipole and quadrupole of NH$_3$ with the quadrupole of the p state of C($^3P$), and the long-range potential $V_{LR}$ varies as a function of $R$ (the distance between the C atom and the center of mass of NH$_3$) as \citep{Zeimen2003}
\begin{equation} \label{eq:long_range}
V_{LR}(R,\theta)= M_{QD}(\theta) {Q({\rm C}) D({\rm NH_3}) \over R^4} + M_{QQ}(\theta) {Q({\rm C}) Q({\rm NH_3}) \over R^5},
\end{equation}
where $Q$(C) and $Q$(NH$_3$) are the quadrupoles of C and NH$_3$, respectively, $D$(NH$_3$) is the dipole moment of NH$_3$, and $M_{QD}(\theta)$ and $M_{QQ}(\theta)$ are 3\,$\times$\,3 matrices that account for the quadrupole-dipole and quadrupole-quadrupole interactions, respectively \citep{Zeimen2003}. In the present treatment we neglect the quadrupole-quadrupole contribution and the dependence on the angle $\theta$. We thus consider an isotropic long range potential $A/R^4$, where $A$\,=\,2\,$M_{QD}$\,$Q$(C)\,$D$(NH$_3$). The parameter $A$ is obtained by fitting to CCSD(T) calculations, with the aVTZ basis set. The ab initio calculations are done for NH$_3$ in its equilibrium geometry, with its symmetry axis along the $z$ axis and its center of mass at origin, varying the $R$\,=\,$z$ coordinate of the C atom. We obtain a value of $A$\,=\,$-$0.85 hartree bohr$^4$. This is similar to the Langevin case, and the capture rate coefficient $k_c$ is then given by
\begin{equation} \label{eq:capture_rate}
k_c(T) = 2\pi \sqrt{A/\mu}~~q_e(T) = 3.13 \times 10^{-10} {\rm cm^3 s^{-1}}~~q_e(T),
\end{equation}
where $\mu$ is the reduced mass of the C + NH$_3$ system and $q_e$ is the electronic partition function of C($^3P$). Assuming that only the 3 lowest spin-orbit states of C($^3P$) yield trapping, we get
\begin{equation}
q_e(T) = {1 + 2 \exp(-23.62/T) \over 1 + 3 \exp(-23.62/T) + 5 \exp(-62.46/T)}.
\end{equation}

The calculated capture rate coefficient is compared in Fig.~\ref{fig:rate} with the experimental reactive rate coefficient measured at different temperatures by \cite{Hickson2015}. We note that the capture rate coefficient is an upper limit to the reactive rate coefficient because once the complex is formed it can either come back to the reactants C + NH$_3$ or evolve to form different products (H$_2$CN, {\textit cis-/trans-}HCNH, or H$_2$NC).

According to the statistical RRKM theory, the micro-canonical rate coefficient $K$ leading to any possible product at a fixed collision energy $E$ and total angular momentum $J$ is expressed as \citep{Marcus1952a,Marcus1952b,Miller1979,Miller1987}
\begin{equation} \label{eq:rrkm_rate}
K(E,J) = {\sum_{v,K} P(E-\epsilon_{JKv}) \over 2 \pi \hbar \rho(E,J)},
\end{equation}
where $\epsilon_{JKv}$ are the energies of the vibrational and rotational levels at the saddle point yielding each product and $\rho(E,J)$ is the density of states of the initial system, which in our case is the intermediate CNH$_3$ (I$_4$). The function $P(E-\epsilon_{JKv})$ in Eq.~(\ref{eq:rrkm_rate}) is the transmission probability over the $\epsilon_{JKv}$ barrier, and can be calculated classically or semiclassically by considering tunneling through a monodimensional Eckart barrier, as proposed by \cite{Miller1979,Miller1987}.

\begin{figure}
\centering
\includegraphics[angle=0,width=0.90\columnwidth]{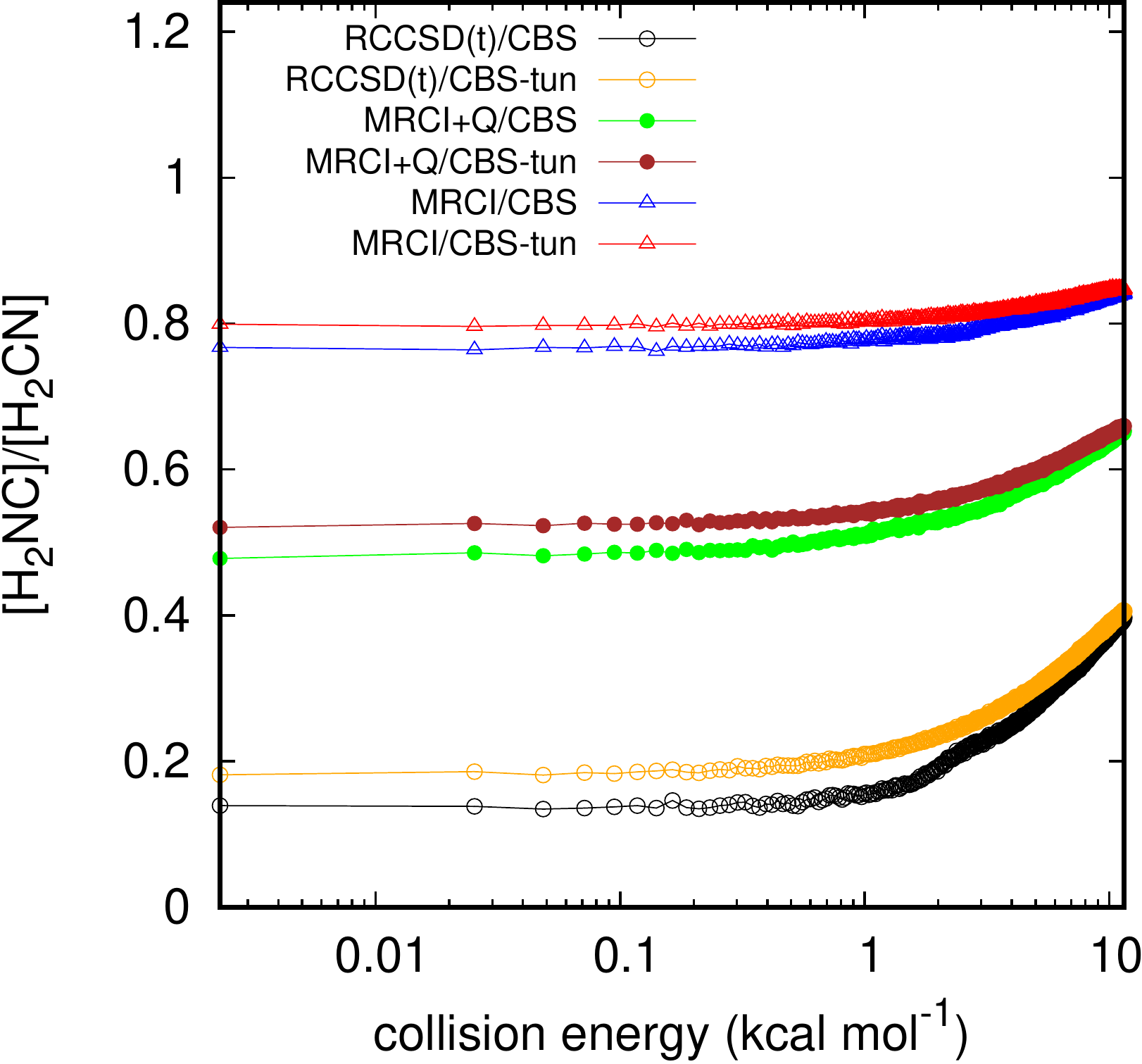}
\caption{Product branching ratio [H$_2$NC]/[H$_2$CN] obtained for the reaction C + NH$_3$ using the VQZ-F12 and CBS results according to Eq.~(\ref{eq:rrkm_rate}), with and without tunneling.} \label{fig:ratio}
\end{figure}

The intermediate CNH$_3$ (I$_4$) will unimolecularly decompose into three channels. The first one is the redissociation back to the reactants C + NH$_3$ (R$_2$) overpassing the barrier of the effective $-A/R^4+\hbar^2\ell(\ell+1)/2\mu R^2$ monodimensional potential (where $\ell$ is the orbital angular momentum of the C atom with respect to the center of mass of NH$_3$), similarly as it is done for the capture model. The second channel leads to the products H + H$_2$NC (P$_4$) overpassing TS$_{\rm I4P4}$ barrier. The third one consists of a statistical mechanism in which TS$_{\rm I3I4}$ is overpassed to reach the I$_3$ intermediate. Along this mechanism, the system will proceed towards I$_2$ and I$_1$ intermediates. This complex-forming mechanism is expected to behave statistically, favoring the exit towards the most stable product, H + H$_2$CN (P$_1$). This was the case calculated by \cite{Cimas2006} in the N + CH$_3$ reaction. Here, we shall assume as a zero-order approximation that once TS$_{\rm I3I4}$ is overpassed the reaction mechanism yields only H + H$_2$CN (P$_1$).

The branching ratio between the two products channels leading to H$_2$NC (P$_4$) and H$_2$CN (P$_1$) is then given as a function of collision energy $E$ as
\begin{equation}
{\left[ {\rm H_2NC} \right] \over \left[ {\rm H_2CN} \right] } (E) = {\sum_{J}(2J+1) K^{{\rm H_2NC}}(E,J) \over \sum_{J}(2J+1) K^{{\rm H_2CN}}(E,J)}.
\end{equation}
The calculated branching ratio [H$_2$NC]/[H$_2$CN] is shown in Fig.~\ref{fig:ratio} as a function of the collision energy using the CBS limit results with the RCCSD(T)-F12, MRCI-F12, and MRCI-F12+Q methods. It is seen that the branching ratio is very dependent on the relative heights of the barriers TS$_{\rm I4P4}$ and TS$_{\rm I3I4}$, which in turn are very dependent on the method, i.e., on how electronic correlation is included. Saddle points correspond to the configurations where bonds are formed and destroyed, and hence, multi-reference methods are in principle more accurate. For this reason, we consider that MRCI-F12 and MRCI-F12+Q results are the most accurate in the present work. The inclusion of tunneling only accounts for corrections of about 2\,\%. It is worth noting that the energy differences between the barriers TS$_{\rm I4P4}$ and TS$_{\rm I3I4}$ involved in this reaction are probably close to the accuracy of state-of-the-art \textit{ab initio} methods. For all explored methods, the [H$_2$NC]/[H$_2$CN] branching ratio is almost constant for collision energies below $\sim$\,0.3 eV. Above this energy, the ratio increases because the density of states at TS$_{\rm I4P4}$ increases faster than at TS$_{\rm I3I4}$ due to the smaller rotational constants. Using the MRCI+Q-CBS results, the rotational constants for TS$_{\rm I4P4}$ are 5.18 cm$^{-1}$, 0.98 cm$^{-1}$, and 0.96 cm$^{-1}$, while for TS$_{\rm I3I4}$ they are 6.49 cm$^{-1}$, 0.87 cm$^{-1}$, and 0.86 cm$^{-1}$.

We can conclude that the [H$_2$NC]/[H$_2$CN] branching ratio at 10 K should have a value between 0.5 and 0.8. This value is to be compared with the abundance ratio H$_2$NC/H$_2$CN observed in cold dense clouds in this work, which lies between 0.51 and $>$\,2.7 (see Table~\ref{table:column_densities}). Taking into account the significant variation of the predicted branching ratio [H$_2$NC]/[H$_2$CN] with the method employed and the spread of H$_2$NC/H$_2$CN ratios found in the targeted clouds, the reaction C + NH$_3$ could perfectly be the main source of H$_2$NC in cold dense clouds, and also in the redshifted galaxy in front of the quasar PKS\,1830$-$211, where the abundance ratio H$_2$NC/H$_2$CN is 0.27.

\section{Concluding remarks}

We carried out an observational survey to search for the metastable isomer H$_2$NC toward various cold dense and diffuse clouds to determine its abundance and understand the chemistry of this recently discovered molecule. Simultaneously the most stable isomer, H$_2$CN, was also searched for in the same sources. We were mainly interested in exploring two particular hypotheses suggested by the discovery of H$_2$NC. First, whether or not the H$_2$NC/H$_2$CN abundance ratio behaves as the HNC/HCN ratio, with values around one in cold dense clouds and below one in diffuse clouds. And second, whether or not H$_2$NC is chemically connected to ammonia.

We detected H$_2$NC in most of the cold dense clouds targeted, which indicates that this species is widespread in this kind of interstellar clouds. On the other hand, we did not detect neither H$_2$NC nor H$_2$CN toward two diffuse clouds which seemed among the most favorable for detection, most likely because of the lower molecular richness of diffuse clouds compared to dense clouds. The non detection of H$_2$NC toward diffuse clouds did not allowed to confirm the hypothesis that the H$_2$NC/H$_2$CN abundance ratio behaves similarly to the HNC/HCN in dense and diffuse clouds. On the other hand, we found that the column density of H$_2$NC is correlated with that of NH$_3$ because the sources where H$_2$NC is most abundant are also very abundant in ammonia. This latter fact suggests that these two molecules are chemically connected, probably because ammonia is a precursor of H$_2$NC through the C + NH$_3$ reaction.

Previous studies of the C + NH$_3$ reaction favored H$_2$CN as the main product. We thus revisited this reaction from a theoretical point of view and found that the two isomers H$_2$CN and H$_2$NC can be formed through two different reaction pathways which involve two different transition states, the energies of which are predicted to be very similar and slightly below that of the reactants. The product branching ratio H$_2$NC/H$_2$CN is predicted to be between 0.5 and 0.8, which is of the order of the abundance ratios H$_2$NC/H$_2$CN observationally found in cold dense clouds, which range from 0.51 to $>$\,2.7. We therefore conclude that the reaction C + NH$_3$ is probably the main formation route to H$_2$NC in interstellar clouds.

\begin{acknowledgements}

We acknowledge funding support from Spanish Ministerio de Ciencia e Innovaci\'on through grants PID2019-106110GB-I00, PID2019-107115GB-C21, PID2019-106235GB-I00, and PID2021-122549NB-C2. We thank the referee for useful comments that helped to improve this article.

\end{acknowledgements}

\end{document}